# A Conversation with Jayaram Sethuraman

Myles Hollander

*Abstract.* Jayaram Sethuraman was born in the town of Hubli in Bombay Province (now Karnataka State) on October 3, 1937. His early years were spent in Hubli and in 1950 his family moved to Madras (now renamed Chennai). He graduated from Madras University in 1957 with a B.Sc. (Hons) degree in statistics and he earned his M.A. degree in statistics from Madras University in 1958. He earned a Ph.D. in statistics from the Indian Statistical Institute in 1962. Before returning to ISI in 1965 as an Associate Professor, he was a Research Associate at the University of North Carolina 1962–1963, at Michigan State University in 1963–1964 and at Stanford University 1964–1965. After three years at the ISI, Sethuraman moved to Florida State University in 1968 as Full Professor. During his career at FSU, he made sojourns as Visiting Professor to the University of Michigan, 1974–1975, the ISI in fall 1977, as a Visiting Professor and Acting Head, ISI Bangalore Center, 1979–1980. He was a senior ASA/NSF/NIST Fellow 1994–1995 and a Fulbright Senior Researcher at ISI Bangalore 1995–1996.

Although Sethuraman officially retired on January 31, 2004 and was named Professor Emeritus at FSU, he continues to be extremely active. He participates in all activities in the Department of Statistics and holds a Courtesy Professor appointment in the Department of Religion. He held an appointment as Professor, University of Pittsburgh in the fall of 2004, and was a Fulbright Senior Lecturer at the Indian Statistical Institute of Technology, Chennai, 2005.

Sethuraman has been a superior researcher throughout his career, making important contributions in many areas including asymptotic distribution theory, large deviations theory, moderate deviations theory for which he was the pioneer, limit theory, nonparametric statistics, Dirichlet processes and Bayesian nonparametrics, stopping times for sequential estimation and testing, order statistics, stochastic majorization, Bahadur and Pitman efficiency, Markov chain Monte Carlo, reliability theory, survival analysis and image analysis. Throughout his career, he has enjoyed continuous external research support from the U.S. Army Office of Research and support from the Academy of Applied Science for mentoring high school students.

Jayaram Sethuraman has received many recognitions for his contributions to the discipline of statistics and to the advancement of science among future scholars. He was elected Fellow of the Institute of Mathematical Statistics (1968) and the American Statistical Association (1971), and became an elected member of the International Statistical Institute (1972). He received the U.S. Army S. S. Wilks Award (1994), was the R. A. Bradley Lecturer, University of Georgia (1995), received the Teaching Incentive Program Award, FSU (1995), and the Professorial Excellence Award, FSU (1996). He was chairman of the FSU Statistics Department 1987–1990. Sethuraman received an ASA Service Award (2001), the President's Continuing Education Award, FSU (2002), and the Bhargavi and C. R. Rao Prize, Pennsylvania State University (2005). In 1993 he was named the Robert O. Lawton Distinguished Professor, FSU. This award is made to only one faculty member per year and is the University's highest faculty honor.

*Myles Hollander is Robert O. Lawton Distinguished Professor and Professor Emeritus, Department of Statistics and Statistical Consulting Center, Florida State University, Tallahassee, Florida 32306-4330, USA e-mail: holland@stat.fsu.edu.*

The following conversation took place in Myles Hollander's office at the Department of Statistics, Florida State University, Tallahassee, on July 7, 2006.

## GROWING UP IN HUBLI AND MADRAS

**Myles:** Sethu, it's a pleasure to be able to have this conversation with you today. Let's begin with your early years in India. You grew up in Hubli.

**Sethu:** Yes, Hubli is about 500 miles northwest from Madras. It's a railroad town. It has a railway workshop, so there are support offices for the railway in Hubli and my father was working in one of those railway offices.

**Myles:** What type of work did he do in that office?

**Sethu:** He was a clerk in the Electrical Engineering Department.

**Myles:** How large is your family?

**Sethu:** We are five surviving brothers. There were eight children born in the family. All boys. However, the first three died within a year or two. So we could have been counted in the famous eight-children data of R. A. Fisher, as a family with all eight boys. I'm the fourth child and the oldest surviving son. I have four younger brothers. As was customary in those days, my mother worked at home showering love and affection on us children. Our grandmother also was living with us.

**Myles:** What did you do as a youngster in Hubli?

**Sethu:** Well, I played and generally enjoyed myself. I went to school in Hubli. I went to St. Mary's High School, a Catholic Jesuit school with teachers who were Jesuit fathers coming from foreign countries—Switzerland, Germany, France and so on. I received a good education at that school. The school enforced a tight discipline and we had to wear uniforms, a pith hat, black shoes, etc., all adding not just a little to the cost of education. We could not afford real bats and wickets to play cricket, and so we played versions of cricket with tennis balls and home-fashioned wooden bats, etc.

**Myles:** What sparked your interest in mathematics?

**Sethu:** Actually, I was doing quite well in the mathematics class and the teacher started to give me all the homework of my classmates to be graded; so I was grading my own class! He would look it over and hand it back to students. It was a good opportunity to stay up on mathematics and do well in school.

**Myles:** How early on did you decide you wanted to pursue math and statistics in college?

**Sethu:** I was not even thinking of college at that time, but somebody told me if you become a teacher you get three months of summer vacation and you get paid during that time (this is true in India) and it is a great job. And so I said to myself, "I should become a teacher." It was only later that I got to know about career possibilities in mathematics and statistics. But going to college itself was also an iffy thing. It just happened, luckily, that I went to college.

**Myles:** You went to Madras.

**Sethu:** I went to Madras and finished high school in Madras. (Madras has been renamed Chennai now.) After that I thought I won't go to college, I'd just take a job and support the family.

**Myles:** You went to high school in Madras before you entered Madras University. What motivated the transition from Hubli to Madras?

**Sethu:** My father moved to Hubli a long time ago and stayed there for almost 20 years. As we grew older, he wanted to get back to Madras because he felt education in high school was better in Madras than in Hubli. After much waiting, he got the transfer to Madras in 1950. I was 13 years old at that time and then joined a high school in Madras. After passing high school I did not think of going to college, but, by another lucky accident, I did enter college.

**Myles:** What was that accident?

**Sethu:** At the end of summer vacation after high school, I went to see the principal of my high school, Sri C. Padmanabha Mudaliyar, just to say hello to him. High schools start earlier than colleges in India. Seeing me, he asked me a question: "Which college are you joining?" I told him that I was not joining any college, I had not even applied to any. He was shocked because I had number one rank in the final public exam from that school, and he said, "No, you had better go to such and such a college and put in an application before you go home."

Because of his insistence I went to that college and filled out a form and told that to my father after I got home. My father was shocked and he said, "We can't afford the fees for you, why did you apply for college?" But then he got convinced and I got admitted to college.

**Myles:** Were there scholarships or did your family have to cover most of the expenses?

**Sethu:** The principal promised that he would be able to get me a scholarship. It was with that assurance that I then went to college. There were exams in the college for a scholarship and I passed those exams and I got a scholarship. Without the scholarship I could not have stayed in college.

**Myles:** Did you live at home while you attended Madras University?

**Sethu:** I did. All the time I was living at home. It was not Madras University. It was one of the constituent colleges of Madras University. Madras University had the authority to conduct common public examinations and grant degrees. You go to any one of several colleges and study for those exams. Madras University holds the exam in the end and gives you the degree.

**Myles:** What is the name of the college?

**Sethu:** Vivekananda College. It was a newly opened college. It was a good college and I really did well and enjoyed mathematics and chemistry, especially. I wanted to pursue chemistry for further studies because I was enamored of all the experiments in chemistry.

**Myles:** You had a good, broad scientific curriculum at this college.

**Sethu:** That's right. You have to specialize when you go to a college and one of the popular specializations in those days was maths, physics and chemistry, which was my option. It was a two-year program at this college, and the degree was called intermediate. After I got my intermediate degree I still had to go to a university to get the B.A. or M.A. I knew that if I continued in the same college and took the math program to get a mathematics B.A. (Hons) degree, I would get a scholarship because I was known very well there, but if I went to another college I might not get a scholarship. It was at this time somebody told me that "If you get a degree in statistics you will get a very high-paying job. A statistician can do the job of 20 people and that is why the salary will be big. It's a new subject, and statisticians are in great demand." I can't remember who it was that said this. It convinced me and so I was excited with statistics, but a statistics B.A. (Hons) was being offered only at the college called Presidency College. It was not clear that I would get a scholarship at Presidency College but I did make an application for the admission and also applied to mathematics at Vivekananda College. Then by a stroke of good luck, a classmate of mine assured me that his father was willing to support me if I did not get a scholarship in statistics at Presidency College. My classmate was also going to join statistics at Presidency College; so on his assurance I joined Presidency College. Eventually I did get a named scholarship called the Sir C. P. Ramaswamy Iyer Scholarship at Presidency College. I thanked my classmate, R. Balakrishnan, and did not use his kind offer.

**Myles:** I know that you graduated with high honors and then you thought as the now-oldest-living son that it was incumbent on you to support your family. You were going to stop your scientific education there, but C. R. Rao intervened. How did this occur?

**Sethu:** I looked for jobs in Madras right after I graduated from statistics. It was very hard to find a job for various reasons, including government policies, about which I will not go into at this time. Finally I got a job at 80 rupees a month as a tutor at Loyola College and my father said, "That's good enough, you can stay here and support us." I almost accepted this job. Meanwhile I had written the entrance exam to go to the Indian Statistical Institute in Calcutta for higher education in statistics. I got an offer from them for a two-year Statistician's Diploma. The stipend was not adequate to support my family. C. R. Rao upgraded that offer to that of a research scholar and still my father said, "Don't go to Calcutta. It's better if you stay in Madras and support the family." I reluctantly agreed with my father and I was going to stay in Madras. That's when, out of the blue, a telegram appeared from C. R. Rao. It was a big shocker to the family and the telegram read, "*Suggest rejecting the job and coming to Calcutta.*"

**Myles:** That was a lightning bolt and it really changed your life, because, if that telegram hadn't arrived, what would have happened?

**Sethu:** I would have taken the tutor job and rotted in Madras. Jobs were not at all available in Madras in those days. Fortunately the telegram came at the correct time. After securing my father's approval, I went to Calcutta.

**Myles:** That telegram came on July 17, 1957. An auspicious day.

**Sethu:** An auspicious day. Yes, it must be an auspicious day. In any case, it changed the course of my life.

## STUDYING AT THE INDIAN STATISTICAL INSTITUTE

**Myles:** You went up to Calcutta and C. R. Rao was the director of the institute, but you wrote your dissertation under Raj Bahadur. How did that happen?

**Sethu:** Dissertations, Ph.D. dissertations, in the Indian Statistical Institute at that time were different. The Indian Statistical Institute did not have authority to grant degrees. You had to register with Calcutta University and get the degree from Calcutta University. At that time at ISI the professors were not guiding students toward Ph.D. dissertations as they do nowadays. A student would pick his own problem and solve that problem halfway or three-fourths of the way and then seek an advisor to finish the dissertation. All of us students would be reading Math Reviews and all sorts of journals to find a problem and then work on the problem. C. R. Rao was the head of the Research and Training School, as the Statistics Division was called at that time. He taught several courses. Seminars were going on. Students gave seminars. Professors gave seminars. You learned a lot from these seminars. You picked up a problem and worked on it on your own and when you had enough stuff, you would go and find some professor willing to advise you. Raj Bahadur was visiting India; he had come back from the U.S. and he was in Calcutta with us. C. R. Rao said, "Why don't you go and meet Raj Bahadur and show your problem to him and see if it is worth a dissertation?"; which is what I did.

**Myles:** I want to hear more about this dissertation under Bahadur because I know that it led to an important paper that deals with convergence of joint distributions and the convergence of the marginals and conditional distributions.

**Sethu:** Yes. When we were looking for problems, Debabrata Basu gave me a paper to read. It was a paper by B. V. Sukhatme, which established the joint asymptotic distribution of the mean and the median, and I was excited with that paper. I was able to prove the same result for the joint distribution of the mean and quantiles by a different technique and that technique led to this idea of establishing the convergence of joint distributions from the convergence of marginals and conditional distributions. That got bigger and I was able to write the whole dissertation and that led to several papers (Sethuraman and Sukhatme, 1959; Sethuraman, 1961a; Sethuraman, 1961b; Sethuraman, 1963).

**Myles:** I know that this idea led to other important applications including the fractile graphical analysis of Mahalanobis and also to fixed interval analysis. Do you want to say something about that?

**Sethu:** Yes. At that time when I was writing my dissertation, P. C. Mahalanobis had introduced the concept of fractile graphical analysis as a new regression method and it was all descriptive and applied. He was challenging the Institute to work on his ideas and supply theoretical support to his methods. I found that my results on convergence of joint distributions based on conditionals and marginals were also applicable to his problem. I found that my result could solve the distributional questions posed by fractile graphical analysis and by fixed interval analysis, another method that I introduced; the dissertation got bigger.

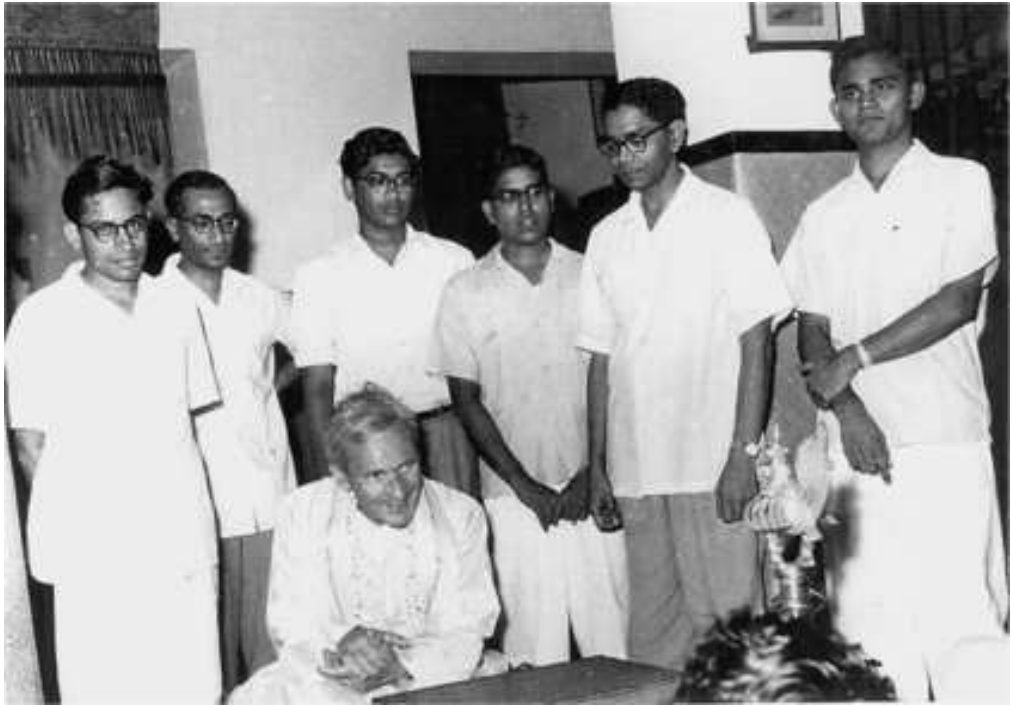

Fig. 1. *Contemporaries at the Indian Statistical Institute: K. R. Parthasarathy, B. P. Adhikari, S. R. S. Varadhan, J. Sethuraman, C. R. Rao, P. K. Pathak (standing), A. N. Kolmogorov (seated), 1961.*

**Myles:** This result about convergence of joint distributions pops up all the time. Over the years it can be found in biometrics, econometrics, reliability papers, nonparametrics. Lots of areas.

**Sethu:** That's true. It has been a little trick that I use in many places to finally prove the result that I am looking for. It appears again and again, including in image analysis, large deviations theory and things like that.

**Myles:** What was the climate at ISI when you were doing your dissertation there? Who were some of the other students that were your contemporaries?

**Sethu:** The leader was V. S. Varadarajan. He had come to ISI a year before me from Statistics at Madras and he inspired us all. He ran seminars. Following him, the rest of us, R. Ranga Rao, K. R. Parthasarathy, S. R. S. Varadhan (winner of the Abel Award), and myself ran our own seminars; we studied function analysis from Dunford and Schwartz and things like that and we got together and discussed many research topics and inspired each other. Nowadays the ISI people are calling it the Golden Age of ISI. By the way, till the time of R. Ranga Rao, people were getting Ph.D.'s from Calcutta University because ISI could not grant degrees. Finally in 1962, the ISI obtained university status and started to grant its own degrees; its first two Ph.D.'s in its first convocation (commencement) were K. R. Parthasarathy and me.

## VISITING APPOINTMENTS AFTER PH.D.

**Myles:** You mentioned image analysis and large deviations theory and it's impressive that throughout your whole career you've been deep but also been amazingly broad. You've had important results in asymptotic distribution theory, large deviations theory, moderate deviations theory and you were really the pioneer of that. You've worked on nonparametric statistics, including sequential nonparametrics work that you did with Richard Savage. Also order statistics, stochastic majorization, Bahadur efficiency work with Herman Rubin, Bayesian nonparametrics, Markov chain Monte Carlo, I could name more. Let's talk about some of these. Let's talk about your work on large deviations.

**Sethu:** After I finished my Ph.D. at the Indian Statistical Institute, C. R. Rao said, "You should go abroad and prove yourself, do research and then you can come back to ISI. I'll give you a three-year leave." He wrote letters for me and I got an offer from Wassily Hoeffding at the University of North Carolina to come for one year. I went to Chapel Hill in 1962 and spent a whole year with him. I decided to work on problems different from my dissertation straight away and I saw this interesting problem on large deviations and that was my first attempt at large deviations. I proved the large deviation result for the empirical distribution function and for families of sample means as they called it at that time. My sojourn for three years continued on to East Lansing for another year and finally Stanford University. During my stay at East Lansing I came in contact with Herman Rubin. When I went to Stanford the next year, Herman was also there on his sabbatical. He and I discussed Bayes risk efficiency and found that moderate deviations was greatly involved in the computation of Bayes risk efficiency and so that's how we introduced moderate deviations (Rubin and Sethuraman, 1965).

**Myles:** Bayes risk efficiency has inspired many others and sometimes it's referred to as Rubin–Sethuraman efficiency.

**Sethu:** Sometimes. Not all the time. That was a very early paper that we wrote, yes, the first paper that we wrote together.

**Myles:** What did you do at Stanford in '64?

**Sethu:** I worked more with large deviations; I taught weak convergence of probability measures and wrote notes for it. I was working with Herman Rubin. I was writing other papers, etc. I was being supported by Herman Chernoff. He invited me several times to his house to taste those famous Chernoff pizzas.

**Myles:** You did later work with Richard when you came to Florida State but you had one theorem with him already, emanating from 1965.

**Sethu:** That's correct.

**Myles:** You were working on sequential nonparametrics while you were at Stanford but Richard Savage was at Florida State, so you must have hooked up somewhere down the line.

**Sethu:** Yes. I got an invitation from Ralph Bradley at Florida State to visit them for a week and to present some talks.

**Myles:** When was that?

**Sethu:** In the Spring of 1965, it might have been March or April. And when I came here and I gave my talks, I guess it was large deviations, I chatted with Richard Savage and I found that he was also working on sequential nonparametrics and I showed him how far I had gone on that problem and where I was stuck. He showed me his progress on the problem. We found that both of us had already, if you combined our solutions, obtained a complete solution to that problem so we were able to show sequential nonparametric likelihood ratio tests based on ranks against Lehmann alternatives terminate with probability 1 (Savage and Sethuraman, 1966).

**Myles:** You had a future potential collaborator in Richard at Florida State and you and he had accomplished a lot in a short time. What did you do next?

**Sethu:** After that one week, I went back to Stanford and then to ISI. Richard and I were very pleased that we were able to accomplish so much within one week. I had fond memories of Tallahassee when I went back to Calcutta in 1965.

**Myles:** What did you do when you went back to Stanford?

**Sethu:** I was working on weak convergence of distributions. I wrote a whole monograph on that and it was supposed to be published. I never published it.

**Myles:** Why not?

**Sethu:** Because Billingsley's book came out at the same time.

**Myles:** Do you still have the notes?

**Sethu:** Yes. I still have the notes.

**Myles:** Do you ever give any thought to modernizing them and writing them up now?

**Sethu:** It can be done and I've given some thought to it.

**Myles:** What happened during the next few years in Calcutta?

**Sethu:** I went back to India in 1965. I went to Calcutta and joined joint duty at ISI, as they called it in those times, as a Reader. Then I took leave and went to see my parents in Madras and then I met Brinda, my future wife, and married her in the same year. Later on we all moved to Calcutta and I was teaching and doing research in Calcutta. Ralph Bradley continued to send me letters saying that he would like me to come to Florida State, he can offer me a position, but I was quite happy in Calcutta and I was not responding to his invitations. But after three years, I found life in Calcutta was more difficult and I thought this was time to get back to the U.S.A. and I came to Florida State in 1968.

## A CAREER AT FLORIDA STATE UNIVERSITY

**Myles:** I was very excited when you came because of professional and personal reasons. I had known you as a friend in 1964 when I was a student at Stanford and now we could renew our friendship and possibly work together. When you returned to Florida State, you and Richard reestablished your collaboration.

**Sethu:** Yes. We wrote a paper on the asymptotic distributions of log-likelihood ratios based on ranks in the two-sample problem that was later published in the Sixth Berkeley Symposium (Savage and Sethuraman, 1972).

**Myles:** That was a coup for Ralph to pursue you and succeed. I know that everyone greatly benefited from it. You were the department's top consultant on research problems and a lot of the faculty benefited from your expertise and assistance. Then three years after you came, Frank Proschan joined us from the Boeing Research Laboratories and you and Frank had a wonderful collaboration.

**Sethu:** Yes, that was a great time. Frank was softspoken and very friendly and would discuss his problems with us very freely, and that openness excited all of us and we got into the mood and did a lot of research. The first work was on Schur functions and majorization and we introduced stochastic majorization in a paper in the *Annals*. This was followed by other problems on stochastic inequalities for distributions (Proschan and Sethuraman, 1977; Nevius, Proschan and Sethuraman, 1977). Finally it led to the DT paper with you, Myles, Myles and Frank (Hollander, Proschan and Sethuraman, 1977).

**Myles:** That DT paper, I continue to see it referenced, but Ingram Olkin and Al Marshal changed the name of the concept from decreasing in transposition to arrangements increasing.

**Sethu:** Right. They call it AI, they didn't like the DT, so people nowadays call it AI.

**Myles:** That paper has a lot of applications for proving inequalities, for example in nonparametric statistics, obtaining inequalities concerning the power of rank tests.

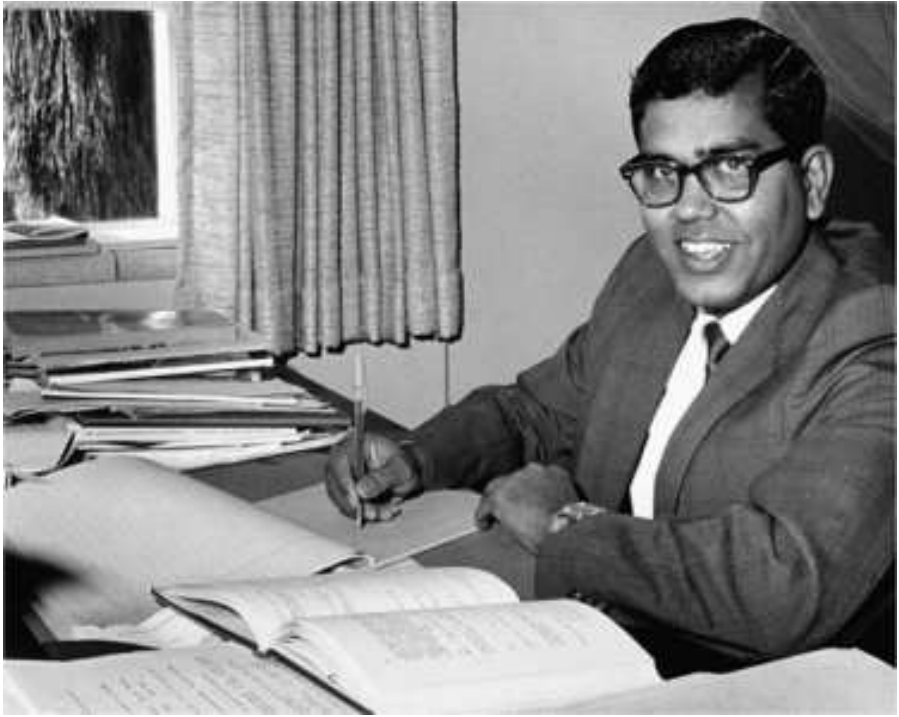

Fig. 2. *Jayaram Sethuraman in his office in the Love Building, Department of Statistics, Florida State University, 1968.*

**Sethu:** That is correct. Those stochastic inequalities are very versatile.

**Myles:** While you were here, in about the early '80s, you had some elegant results on Ferguson's Dirichlet process (Ferguson, 1973) including your construction of the Dirichlet process (Sethuraman and Tiwari, 1982; Sethuraman 1994).

**Sethu:** Yes. In 1979, I believe, David Blackwell was visiting us for one semester from Berkeley and that was the time I was giving a seminar series to the students on Dirichlet processes. Every day I was discovering new results and presenting them in class. One fine day this new representation of the Dirichlet process popped up and Blackwell was one of the first persons to hear this from me and he also supplied a key proof for identifying a distribution from a distributional equation which appears in my paper which was finally published in 1994 (Sethuraman, 1994).

**Myles:** Your representation is used in obtaining many results and continues to be referenced. One of the nice things was that it provided a different approach to proofs and some were easier. For example, the proof that the Dirichlet process concentrates on discrete distributions comes easily.

**Sethu:** It is there in the definition itself. My definition of a Dirichlet process is a random discrete distribution; therefore Dirichlet processes concentrate on discrete distributions.

**Myles:** There was some work to do because you still had to show that your process was the same as Ferguson's Dirichlet process.

**Sethu:** Yes. My representation was concentrated on discrete distributions, but you had to show the finite-dimensional distributions were finite-dimensional Dirichlet. It turns out that you can write down a simple functional equation for the distribution and show that the Dirichlet distribution is the only so-lution to that functional equation. That's a fixed point theorem and we proved that. The same fixed point theorem can be used to prove that the posterior distribution is also Dirichlet. So the complete picture comes out. But more than that, the beauty of the representation is it uses only independent random variables; so all the computations are standard and easy to do. In modern times it's also useful for computation. You can actually obtain samples of the Dirichlet process because of this representation. The other representations cannot help you to do that.

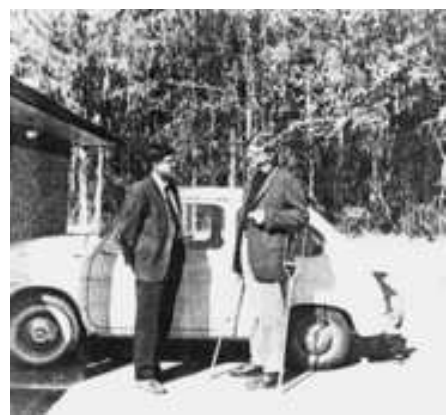

Fig. 3. *Jayaram Sethuraman and Richard Savage in front of Jo Ann and Richard Savage's home, Tallahassee, 1970.*

**Myles:** In 1983 you were the principal speaker at a one-week conference and lecture series on Dirichlet processes at Penn State. How did that arise?

**Sethu:** At Penn State, Jim Lynch arranged for me to come and give this one-week SIAM lecture series and it was on Dirichlet processes. Really, not just Dirichlet, but all sorts of Bayesian nonparametrics. I introduced more families of nonparametric prior distributions which include the Dirichlet process and other processes, nowadays called Pólya tree processes, that can concentrate on continuous distributions or singular distributions, and how to compute posteriors in all cases. This was a long series, a detailed series, some of which have now reappeared in different languages by other people in different notation.

**Myles:** It's also true that Bruce Lindsey's handwritten notes of your lectures at Penn State have been widely circulated. Have you thought of writing a monograph on Bayesian nonparametrics?

**Sethu:** Well, Lindsey's notes were widely circulated and J. K. Ghosh and R. V. Ramamoorthi have included many of the new results from those lectures in their book (Ghosh and Ramamoorthi, 2003).

**Myles:** In the Dirichlet process work one of the interesting and surprising things was that in a sense you showed that the commonly held notion that the $\alpha(R)$ parameter could be interpreted as a prior sample size had problems. You pointed out that as $\alpha(R)$ went to zero, a very strange thing happens; the process concentrates on a degenerative distribution, which is not appropriate for a prior.

**Sethu:** That's correct. The literature likes to refer to the parameter $\alpha(R)$ as a prior sample size and to equate putting $\alpha(R)$ equal to zero as equivalent to no information, but I showed that if you allow $\alpha(R)$ to go to zero and keep the normalized probability measure constant, you'll be getting a very informative prior, namely a degenerate distribution at some random point, and so therefore $\alpha(R)$ going to zero is not the same as saying prior information is zero.

**Myles:** Emad El-Nweihi, Jim Lynch and Chaganty Rao are three of your 23 Ph.D. students with whom you've continued to collaborate on problems extending beyond their dissertations. Would you comment about those results?

**Sethu:** Yes, I've been interacting with Emad El-Neweihi, Jim Lynch and Chaganty Rao after their dissertations were written and published. We published on topics different from their respective dissertations.

With Emad, I published a series of papers in reliability, drawing inspiration from Frank Proschan. They built upon urn models, order statistics and the role and optimal allocation of components in coherent systems (El-Neweihi, Proschan and Sethuraman, 1978, 1986; El-Neweihi and Sethuraman, 1991).

With Jim, I proved a very interesting result of large deviations for processes of independent increments whose moment generating function was not finite for all $t$; the exponential distribution is an example. Previous results assumed that the moment generating function existed for all $t$, but if it exists only on an interval of $t$, results were not available and this was the first paper that obtained the large deviation results for that case. In the process, we showed the parallelism between large deviations and weak convergence and those techniques are useful in the proof of those results (Lynch and Sethuraman, 1987).

With Changanty Rao, I proved results that strong large deviations and local limit theorems can be merged together—one helps in the proof of the other, and so this was a paper in a different direction, not just large deviations, strong large deviation theorems (Chaganty and Sethuraman, 1993).

**Myles:** You, Krishna Athreya and Hani Doss have an important result on convergence in Gibbs sampling.

**Sethu:** Yes, I was excited when I saw the Markov chain Gibbs sampler and I wasn't convinced at that time that conditional distributions determined the joint distribution. In fact, it's not true, but under most circumstances it is. To prove that the Markov chain in the Gibbs sampler actually converges, people are referring to more and more conditions from different books and papers so it was not clear what the conditions were. We studied this problem and we had an elegant result on the convergence of Markov chains in general, and in particular for the Gibbs sampler. A joint paper with Athreya and Doss appeared in the *Annals of Statistics* (Athreya, Doss and Sethuraman, 1996) and it has a more verifiable condition which you can verify especially when one of the processes there is a Dirichlet process. Hani Doss illustrates this in his paper on Gibbs sampling with incomplete data and Dirichlet mixture priors (Doss, 1994).

**Myles:** In the last 15 years you've written a sequence of papers on repair models in the area of reliability. How did this evolve?

**Sethu:** Brown and Proschan introduced a model called imperfect repair (Brown and Proschan, 1983) in which either a perfect repair or a minimal repair was done after each failure. That paper was extended by Block, Borges and Savits (Block, Borges and Savits, 1985). They were proving only distributional results. Frank Samaniego and Lyn Whitaker (Whitaker and Samaniego, 1989) came up with an estimate of the distribution to the first failure as a nonparametric maximum likelihood estimate. They established its properties and we did the same thing using counting processes in much more generality and obtained estimates of the distribution to first failure in a repair model which was the Block–Borges–Savits model (Hollander, Presnell and Sethuraman, 1992). We also obtained confidence bands for that distribution. This was followed by a much more general repair model (Dorado, Hollander and Sethuraman, 1997). The key here was that an observation under repair can be viewed like a censoring, and ideas of censored models can be used. It was a nice use of censoring ideas in a different context. This work is only frequentist in nature and required some peculiar assumptions to be made about the observation period. The sampling scheme was to observe $n$ Block–Borges–Savits processes, each until the time of its perfect repair, but stop at the first failure age such when there is only one process which has not yet experienced a perfect repair. More recently this has been followed by Bayesian methods, using Dirichlet processes and other general processes we called partition-based prior distributions (Sethuraman and Hollander, 2008). This Bayes approach does not require any of the stopping assumptions of the frequentist methods and it is also available under the very general repair models introduced by us.

**Myles:** How did you get started on research with Ulf Grenander?

**Sethu:** We had a new institute established on campus, the Supercomputer Computations Research Institute. We received some money from that institute and we were able to bring in many visitors and one of the visitors was Ulf Grenander. He visited us for a whole week and gave a series of lectures. That got me excited in this area. Grenander was working on a problem at that time; we discussed it a little. Within a few weeks we both had a solution to the problem and so a paper came out which was published after many years, because it just lay idle on our desks. It is called "Mixed Limit Theorems for Pattern Analysis" (Grenander and Sethuraman, 1994), but I also had Ph.D. students work on similar problems; Kurien was one of those students. The papers with Kurien appeared even before my joint paper with Grenander (Kurien and Sethuraman, 1993a, 1993b).

**Myles:** You've taught image analysis to our students.

**Sethu:** Yes, I've run a special topics course in image analysis for several years in the department and there are still lots of problems in that area that I want to work on.

**Myles:** Another person you've worked with is Nozer Singpurwalla. In fact, Nozer and you have the same birthplace. You're both from Hubli and I understand C. R. Rao was born not too far from there.

**Sethu:** That's correct. We found that out only recently when we visited Penn State. C. R. Rao also comes from a town close to Hubli. I met Nozer for the first time when he came to attend a workshop on reliability that Proschan conducted at FSU. It is then we found out that we were born in the same place. We found that we were working on similar problems. Nozer later visited our department when Dennis Lindley was giving his lectures on Bayes methods. It was at this time we discussed accelerated failure testing and wrote two papers (Sethuraman and Singpurwalla, 1981, 1982).

**Myles:** That visit to Penn State—you're not going to bring it up, but I will. It was for the Bharghavi and C. R. Rao prize which you just received in 2005. Thus far, only two people have received that prize, Bradley Efron who was the first recipient, and yourself. It's going to be a distinguished list. I had the pleasure of attending that celebration of you at Penn State and it was wonderful seeing C. R. Rao in action with you at this conference.

**Sethu:** Yes. C. R. Rao honored me by going to great lengths to actually bring back some photographs from way back at ISI, which I have not seen in a long time, and he put them on display at the function to bestow me the award.

**Myles:** I remember Varadhan was in the photograph, and who were some others?

**Sethu:** Yes, that old gang was Parthasarathy, Varadhan, etc. There is also A. N. Kolmogorov in that picture. Kolmogorov was visiting the ISI at that time. We research scholars flocked to his seminars to learn his new research. Kolmogorov could not speak much English, but was fluent in French and German. We had a graduate student from the U.S. called Thomas Weisskopf, who was fluent in those languages and he acted as our official translator.

## FULBRIGHT AWARDS

**Myles:** Let's talk about your Fulbrights. You went to ISI Bangalore in 1995–1996 on a Fulbright and just recently in 2005 you went to IIT in Madras on another Fulbright.

**Sethu:** I've been lucky to have received two Fulbright awards 10 years apart. I was lucky in getting these awards to go to India, both for professional and personal reasons. The first time was at ISI Bangalore. It was a good choice. Statistics is always a strong subject at ISI and I gave a series of lectures on Bayesian nonparametrics. I also worked on image analysis. The second time I went to IIT in Madras hoping to give lectures on image analysis and to attract their students to come to Florida State. But this time I was not so successful. The mathematics department at IIT did not have a strong interest in statistics. So after giving a few talks at IIT, I went over to the Statistics Department at Madras University and gave my lectures there. I was not successful in attracting the IIT students. Their training took them to fields other than statistics for higher education.

**Myles:** Students in India are turning to other fields like computer science rather than statistics. Students from India were a great source for statistics departments over the years, but now many opt for a different career.

**Sethu:** Exactly. Exactly. Students from India nowadays are typically not going into hard sciences or mathematics and statistics. They are going to more "lucrative" fields like business administration or computer sciences. I shouldn't call them "lucrative" fields, but they are in other fields also and they are doing well in those fields, with some of them really excelling in those fields.

## TRAINING HIGH SCHOOL SENIORS

**Myles:** You have a remarkable style of teaching. You walk into class without any notes. It seems like you have it all in your head.

**Sethu:** Yes, I don't need notes. I can make students co-discover with me the topics that I want to teach. Thus the students relate to the topic just as they are being taught. I can gauge the understanding of the students and change my lectures to suit the situation.

**Myles:** You not only teach our graduate students and our undergraduates, but you teach senior high school students in this REAP program that you've been doing since 1981.

**Sethu:** I've been doing it for more than 25 years. It's called the Research in Engineering and Sciences Apprenticeship Program (REAP). It's a nice program that the Academy of Applied Sciences supports with financing from the Army Research Office. I get to pick some high school students, usually high school seniors, and work with them in the summer time. After some initial reading, the students take up a project, finish it, and write a report on the project; this has been going for 25 years so there have been more than 75 students in this program, so far.

**Myles:** You have received the Florida State University President's Continuing Education Award for that program in 2003 and you are still doing it. In fact, just before we started this conversation I found you talking with your REAP students in an office.

**Sethu:** I currently have three REAP students and I need to go and talk to them every day. It was a nice award. Nice to get the award from the president, nice to get the recognition for this extracurricular award, nice to work with high school students and to reach out to the society outside the university.

**Myles:** Since you don't have REAP students take formal classes, tell us about your day-to-day interactions with them, your style of stimulating them.

**Sethu:** The style is similar to my teaching style in my classes. Here the students are closer to me and listen to me more intently. When they understand something or discover something you can see their eyes light up. I tell them that is their "Aha" moment. Instead of answering their questions directly, I pose more related questions which lead them to the solution of the first question. Generally the students learn elementary probability and work on a project that uses probability with some rudimentary statistics. I find new projects by looking at magazines like Chance and books on elementary statistics.

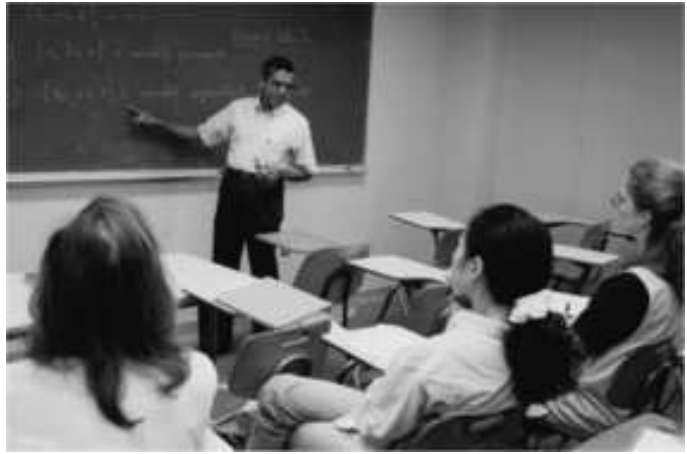

Fig. 4. *Jayaram Sethuraman teaching a class, 1993.*

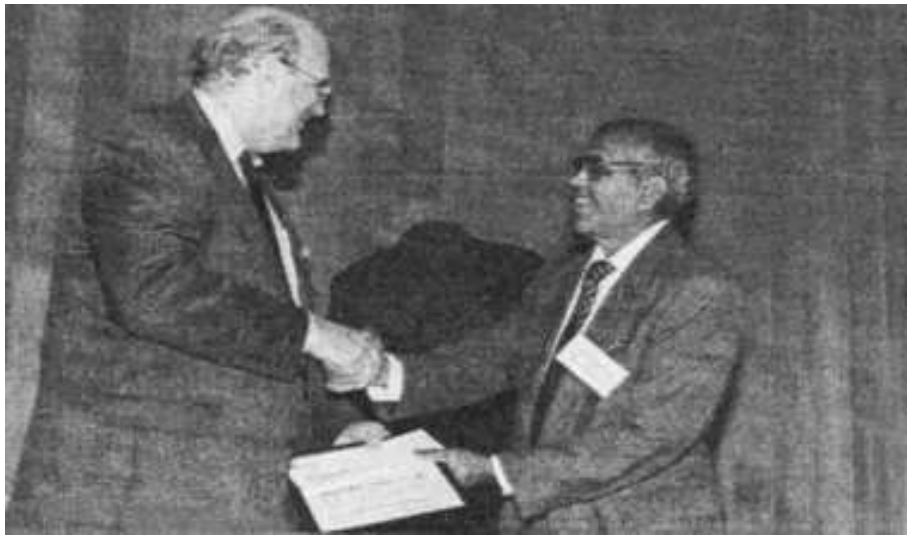

Fig. 5. *Jayaram Sethuraman receiving the President's Continuing Education Award from FSU Provost Larry Abele, 2002.*

**Myles:** I've seen the REAP students' letters of praise about you. Many of them go on to interesting careers in science.

**Sethu:** Right. Quite a few have gone on into statistics, or actuarial sciences, or mathematics.

### FAMILY LIFE AND HOBBIES

**Myles:** You came here in 1968 with your wife Brinda, although I think you had already been married for 3 years. Tell me about your family.

**Sethu:** My wife Brinda and I have been married for 41 years. We have a son, Sunder, who is at Iowa State University and a daughter, Nitya, who is at the University of Indiana and they are pursuing their academic careers. We also have two grandchildren Anupama (a girl) and Adithya (a boy).

**Myles:** I could say that both apples fell pretty close to the tree in the sense that Sunder is a mathematician/probabilist and Nitya is in linguistics and cognitive science. You're a linguistic scholar.

**Sethu:** Yes. I have an interest in Sanskrit, especially. Right from childhood I wanted to study more Sanskrit, but statistics is a higher-paying field and statistics has been my profession.

**Myles:** You currently have a courtesy appointment in our Department of Religion and you're teaching Sanskrit to students there.

**Sethu:** That's correct. I've taught Sanskrit to students over the years in the Religion Department; so as recognition, they have given me a courtesy professorship appointment and I'm willing to teach Sanskrit to any student who wants to come and learn.

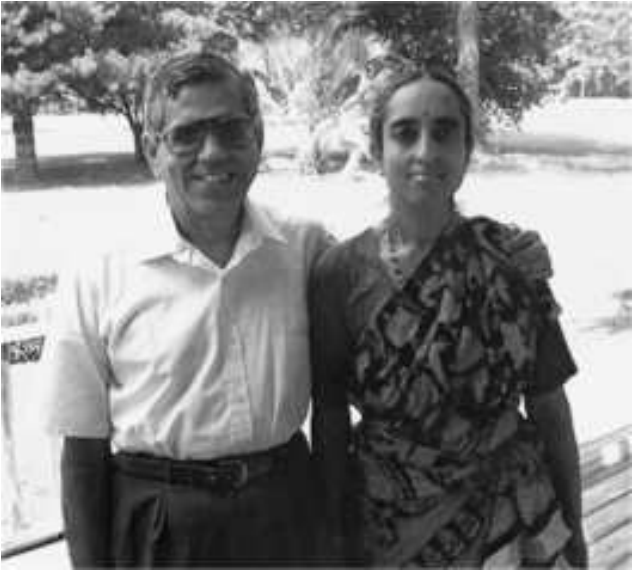

FIG. 7. *Jayaram and Brinda Sethuraman at the home of Glee and Myles Hollander, 2004.*

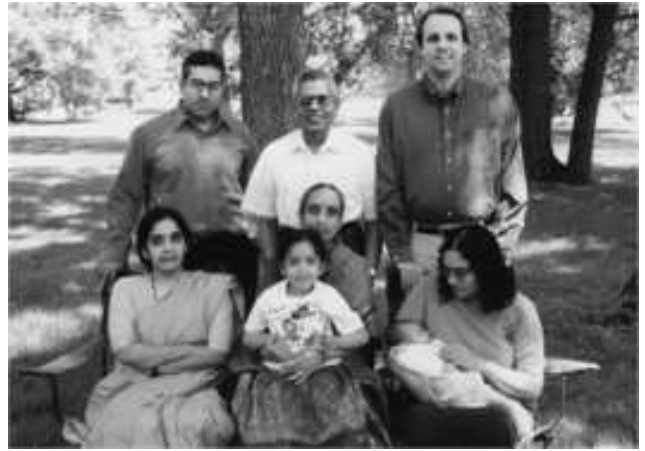

FIG. 8. *Sunder Sethuraman, Jayaram Sethuraman, Aarre Laakso (standing), Lalitha Madhavan, Brinda Sethuraman with Anupama, Nitya Sethuraman with Adithya (seated), Bloomington, Indiana, 2005.*

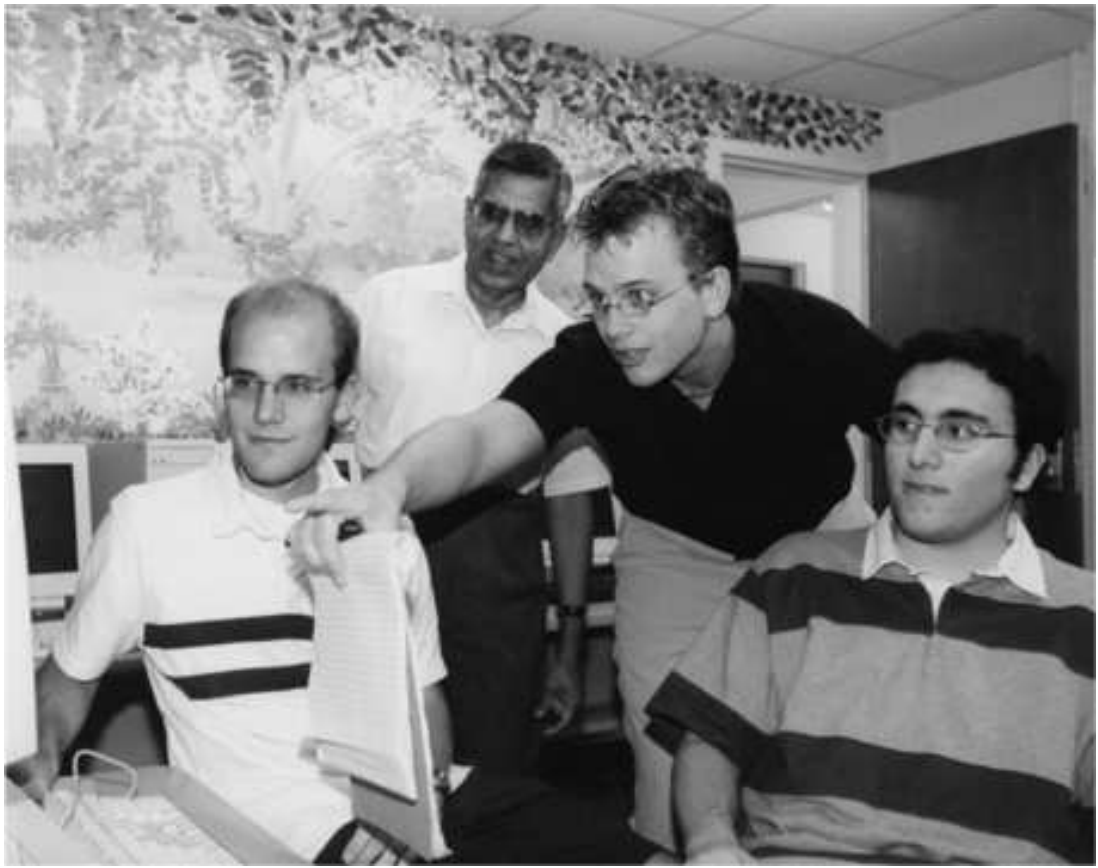

FIG. 6. *Jayaram Sethuraman with a group of REAP students, 2003.*

I'm doing that with some students from our own department right now.

**Myles:** Have you traveled elsewhere to talk with Sanskrit scholars? I seem to remember you communicating with a man in California.

**Sethu:** I went to Montreal, Canada, recently because I got an invitation to come and speak at a Sanskrit conference and I read them some of my compositions, one of which was a political satire. It was well appreciated at that meeting. There was a person from California who sent a response to one of my electronic postings of a Sanskrit verse. I got in touch with him by email and found that he was in San Diego where our daughter was in graduate school. So I finally was able to meet this person, C. V. Mahalingam. He was a senior person who worked all day on his hobbies—Sanskrit and music. We got close and I benefited from his advice on many questions in Sanskrit. I felt very sorry when he passed away a few years ago.

**Myles:** What are some of the topics of your Sanskrit poems?

**Sethu:** I've written two Sanskrit poems which are available on my website page. One is devoted to Lord Venkateswara. There are many Hindu temples in the U.S.A. with a sanctum for Lord Venkateswara and many of these places are also described in this poem. This poem has been reviewed by Vasudha Narayanan, a professor of religion at the University of Florida. The review forms a part of her paper on Hindu observances in the U.S. A similar review has appeared in an Indian newspaper in the U.S. Another of my works is a biographical poem which mentions many events in my life, and it's a praise poem to the goddess, who I call the goddess of infinite mercy. I have also started to write a fictional poem on the Titanic and have written so far only two cantos. All these are available on my webpage. I update them whenever I add new verses. Since I use LaTeX to type Sanskrit this is easy to do; for instance, I do not have to renumber the verses.

During my Fulbright year last year I met an Indian professional musician. She liked my Sanskrit poems and she has sung them and put them on a CD. I am thinking of putting her recording on the webpage.

**Myles:** You and Brinda are avid gardeners. Glee and I have enjoyed many tasty fruits and vegetables from your garden.

**Sethu:** There are many tropical vegetables that are not available in supermarkets, though they are now becoming available in ethnic stores. The weather in Tallahassee is excellent for growing these vegetables. Brinda and I make the best of the local conditions to have a nice garden to grow many tropical vegetables and lots of flowers. Seeds for these vegetables are available in local seed catalogs.

**Myles:** What is your opinion of our field which has proved so challenging and exciting over the years. Where are we going?

**Sethu:** I think that the future of statistics is very bright. It's going on to more and more challenging and difficult problems. Most of our previous work looks very limited in this respect. The future is unlimited. There are no neat closed-form solutions for many of the modern problems. Heavy computation and deep intuition are required to solve the new problems in statistics. I don't know how much I will contribute to the future of statistics in this direction, but I'm still keeping myself abreast of the new things that are coming along and working on others.

**Myles:** Sethu, you and I have been friends for 42 years since we first met at Stanford. It's been a tremendous pleasure and privilege for me to be able to have this conversation with you. I know you've helped many people in your career and I'm grateful for all the help and companionship you've given me.

**Sethu:** I appreciate this conversation very much, Myles, and our long friendship. Brinda and your wife Glee are good friends and the two families have been getting along well all the time. We have our common joint work and common interests and we will continue to be great friends. I'm pleased that we had this conversation and will cherish it for a long time.

## ACKNOWLEDGMENT

Jayaram Sethuraman and Myles Hollander thank Pamela McGhee for transcribing this conversation.